\documentclass[a4paper, 10 pt, conference]{IEEEtran}
\pdfoutput=1

\usepackage[utf8]{inputenc}
\usepackage[T1]{fontenc}
\usepackage{graphicx}
\usepackage{breqn}
\usepackage{subcaption}
\usepackage{amsmath}
\usepackage{amsthm}
\usepackage{url}
\usepackage{footnote}

\usepackage[ruled,vlined,linesnumbered]{algorithm2e}

\usepackage[normalem]{ulem}

\newtheorem{theorem}{Theorem}
\numberwithin{theorem}{subsection}
\newtheorem{proposition}[theorem]{Proposition}
\newtheorem{corollary}[theorem]{Corollary}

\newtheorem{definition}[theorem]{Definition}
\newtheorem{remark}[theorem]{Remark}

\usepackage{stackengine}
\newcommand\underparen[1]{\@ifnextchar_{\uphelp{\uparen{#1}}}{\uparen{#1}}}
\makeatother
\def\uphelp#1_#2{\ensurestackMath{\stackunder[1pt]{#1}{\scriptstyle #2}}}
\newcommand\uparen[1]{\setbox0=\hbox{$#1$}\ensurestackMath{%
  \stackunder[0pt]{#1}{\rotatebox{90}{$\left(%
  \rule[\dimexpr-.5\wd0+\dp\strutbox-1.3pt]{0pt}{\wd0}\right.$}}%
}}
\usepackage{amssymb} % for triangleq
\newcommand{\defeq}[0]{\triangleq}
\usepackage{xspace}

\newcommand{\keepcomment}{1} % 1 - Keep comments, 0 - Hide comments

\ifnum\keepcomment=1
	\usepackage[colorinlistoftodos,textsize=scriptsize]{todonotes} % todo
    \newcommand{\stkout}[1]{\ifmmode\text{\sout{\ensuremath{#1}}}\else\sout{#1}\fi}
    
\else
	
	\usepackage[disable]{todonotes} % todo
\fi

\usepackage{paralist} % To use compactitem

\usepackage{textcomp} % to use €

\newcommand{\cotwo}{CO\textsubscript{2}}
\usepackage{siunitx} % To use \SI
\usepackage{booktabs} % To use \toprule
\newcommand\Alt{j}
\newcommand\Altt{{j'}}
\newcommand\Ind{i}

\newcommand\SocUtSym{b}
\newcommand\SocWelfareSym{B}

\newcommand\IndUtSym{V}
\newcommand\IndUtAfterSym{U}
\newcommand\IncSym{y} % incentive

\newcommand\TrSym{y} % transfer
\newcommand\TotIncSym{Y}
\newcommand\WeightSym{w} % incentive

\newcommand\NumOfInd{m}
\newcommand\SplitInd{s}
\newcommand\SplitAlt{t}

\newcommand\DecSym{x}

\newcommand\Budget{Q}
\newcommand\BudgetUsed{\tilde{\Budget}}

\newcommand\NonDomSetSym{\mathcal{R}}

\newcommand\EffSym{e}

\newcommand\Curve{\mathcal{C}}

\newcommand\IndSet{\mathcal{I}}

\newcommand\ProbSym{\mathbb{P}}
\newcommand\ExpSym{\mathbb{E}}

\newcommand\AccProbSym{\pi}

\newcommand\InitSetSym{\mathcal{N} }

\newcommand\OfInd[1]{_{#1}}

\newcommand\OfIndAlt[2]{_{#1,#2}}
\newcommand\OfIndAltDef{\OfIndAlt{\Ind}{\Alt}}

\newcommand\InitSet[1]{\InitSetSym\OfInd{#1}}
\newcommand\InitSetDef{\InitSet{\Ind}}
\newcommand\IndUt[2]{\IndUtSym\OfIndAlt{#1}{#2}}
\newcommand\IndUtDef{\IndUtSym\OfIndAltDef}

\newcommand\IndUtAfterFunc[3]{\IndUtAfterSym\OfIndAlt{#1}{#2}}

\newcommand\IndUtAfterFuncDef{\IndUtAfterFunc{\Ind}{\Alt}{\Policy}}
\newcommand\SocUt[2]{\SocUtSym\OfIndAlt{#1}{#2}}
\newcommand\SocUtDef{\SocUtSym\OfIndAltDef}
\newcommand\Inc[2]{\IncSym\OfIndAlt{#1}{#2}}
\newcommand\IncDef{\IncSym\OfIndAltDef}

\newcommand\TrDef{\TrSym\OfIndAltDef}
\newcommand\IncPolicy{\mathbf{\IncSym}}

\newcommand\Policy{\mathbf{\TrSym}}
\newcommand\Weight[2]{\WeightSym\OfIndAlt{#1}{#2}}

\newcommand\WeightDef{\WeightSym\OfIndAltDef}

\newcommand\AltPair[2]{[#1,#2]}
\newcommand\AltPairDef{\AltPair{\Ind}{\Alt}}
\newcommand\DefAlt[1]{\Alt^0\OfInd{#1}} % default alternative
\newcommand\DefAltDef{\DefAlt{\Ind}}

\newcommand\Dec[2]{\DecSym\OfIndAlt{#1}{#2}} % decision variable
\newcommand\DecDef{\Dec{\Ind}{\Alt}}
\newcommand\MaxSocUt[1]{\SocWelfareSym^*(#1)}
\newcommand\MaxSocUtDef{\MaxSocUt{\Budget}}

\newcommand\AlgSocUt[1]{\SocWelfareSym(#1)}
\newcommand\AlgSocUtDef{\AlgSocUt{\TotIncSym}}

\newcommand\NonDomSet[1]{\NonDomSetSym\OfInd{#1}}
\newcommand\NonDomSetDef{\NonDomSet{\Ind}}
\newcommand\NumOfNonDom[1]{|\NonDomSet{#1}|}
\newcommand\NumOfNonDomDef{\NumOfNonDom{\Ind}}
\newcommand\IncrSocUt[2]{\tilde{\SocUtSym}\OfIndAlt{#1}{#2}}
\newcommand\IncrSocUtDef{\IncrSocUt{\Ind}{\Alt}}
\newcommand\IncrWeight[2]{\tilde{\WeightSym}\OfIndAlt{#1}{#2}}
\newcommand\IncrWeightDef{\IncrWeight{\Ind}{\Alt}}
\newcommand\IncrEffSym{\tilde{\EffSym}}
\newcommand\Eff[2]{\EffSym\OfIndAlt{#1}{#2}}
\newcommand\EffDef{\Eff{\Ind}{\Alt} }
\newcommand\IncrEff[2]{\IncrEffSym\OfIndAlt{#1}{#2}}
\newcommand\IncrEffDef{\IncrEff{\Ind}{\Alt}}
\newcommand\TotInc{\TotIncSym}

\newcommand\TotNonDomSet{\NonDomSetSym}
\newcommand\TotSocUt{{\SocWelfareSym}}

\newcommand\NumOfAlt[1]{|\InitSet{#1}|}
\newcommand\NumOfAltDef{\NumOfAlt{\Ind}}
\newcommand\NumOfConvHull[1]{|\NonDomSet{#1}|}
\newcommand\NumOfConvHullDef{\NumOfConvHull{\Ind}}
\newcommand\TotNumOfConvHull{|\TotNonDomSet|}

\newcommand\RandomSym{\epsilon}
\newcommand\Random[2]{\RandomSym\OfIndAlt{#1}{#2}} % random term
\newcommand\RandomDef{\Random{\Ind}{\Alt}}
\newcommand\AltEq[2]{\Alt^{*}_{#1}} % equilibrium alternative
\newcommand\AltEqDef{\AltEq{\Ind}{\Policy}}
\newcommand\IndUtDet[2]{\hat{\IndUtSym}\OfIndAlt{#1}{#2}}
\newcommand\IndUtDetDef{\IndUtDet{\Ind}{\Alt}}
\newcommand\IncDet[2]{\hat{\IncSym}\OfIndAlt{#1}{#2}}
\newcommand\IncDetDef{\IncDet{\Ind}{\Alt}}

\newcommand\AccProb[3]{\AccProbSym_{#1, #2}(#3)}
\newcommand\AccProbDef{\AccProb{\Ind}{\Alt}{\IncDef}}

\newcommand\IterSym{k}
\newcommand\OfIter[1]{^{[#1]}}
\newcommand\OfIterDef{\OfIter{\IterSym}}

\DeclareMathOperator*{\argmax}{arg\,max}

\title{
Large-Scale Allocation of Personalized Incentives
}

\author{
    Lucas Javaudin$^1$, Andrea Araldo$^2$, André de Palma$^1$\\
    $^1$CY Cergy University; lucas.javaudin@cyu.fr; andre.de-palma.cyu.fr\\
    $^2$Télécom SudParis, Institut Polytechnique de Paris; andrea.araldo@telecom-sudparis.eu
}

\usepackage{array}
\newcolumntype{C}[1]{>{\centering\arraybackslash}m{#1}}
\usepackage{pgfplots}
\pgfplotsset{width=8cm,compat=1.9}

\begin{document}

\maketitle
\thispagestyle{plain}
\pagestyle{plain}

%%%%%%%%%%%%%%%%%%%%%%%%%%%%%%%%%%%%%%%%%%%%%%
%%%%%%%%%%%%%%%%%%%%%%% ABSTRACT %%%%%%%%%%%%%
%%%%%%%%%%%%%%%%%%%%%%%%%%%%%%%%%%%%%%%%%%%%%%
\begin{abstract}
 We consider a regulator willing to drive individual choices towards increasing social welfare by providing incentives to a large population of individuals.

    For that purpose, we formalize and solve the problem of finding an optimal personalized-incentive policy: \emph{optimal} in the sense that it maximizes social welfare under an incentive budget constraint, \emph{personalized} in the sense that the incentives proposed depend on the alternatives available to each individual, as well as her preferences.
    We propose a polynomial time approximation algorithm that computes a policy within few seconds and we analytically prove that it is boundedly close to the optimum.
    We then extend the problem to efficiently calculate the Maximum Social Welfare Curve, which gives the maximum social welfare achievable for a range of incentive budgets (not just one value).
    This curve is a valuable practical tool for the regulator to determine the right incentive budget to invest.

    Finally, we simulate a large-scale application to mode choice in a French department (about 200 thousands individuals) and illustrate the effectiveness of the proposed personalized-incentive policy in reducing \cotwo{} emissions.
\end{abstract}

\maketitle

\section{Introduction}

%We consider a population of individuals, each of them facing a specific choice set.
%The individuals choose their preferred alternative which, does not correspond necessarily to the socially optimal choice.
%In this section, we present the mathematical formulation of the problem used throughout this paper.
%The framework that we propose could potentially be adapted to many real world situations in transportation area, in marketing and in several other fields, in particular, where externalities are prevalent.

Taxes and subsidies in transportation are often perceived by the population as unfair, since they neglect the alternatives actually available to each individual and the individual preferences.
On the other hand, with the increase in information available to governments \cite{clarke2014governments}, economic policies can be improved to consider the peculiarities of each individual.
We propose a policy of personalized incentives in a framework where individuals choose between multiple alternatives options.
A regulator has a limited budget that he can use to propose monetary incentives, with the goal to induce individuals to change their choice toward socially-better ones. Most incentive policies in the literature are not personalized~\cite{Sun2020,HaiYang2020}, with some exception~\cite{Araldo2019}. Differently from the latter, we seek for a method with provable performance bounds.

We define the optimal personalized-incentive policy as the allocation of incentives that maximizes social welfare (defined as the reduction of \cotwo{} emissions in the example above), for a given budget.
We formalize the problem of finding a personalized-incentive policy maximizing social welfare under the regulator's budget constraint and show that it reduces to the well-known Multiple-Choice Knapsack Problem (MCKP -- \S\ref{sec:policy}), which has been used in several contexts, like Economics~\cite{Colorni2017} and Computer Science~\cite{AraldoSAC2020}.
To approximate the optimal policy in polynomial time, we adapt a greedy algorithm from the Operations Research literature and we analyze some of its analytical (e.g., suboptimality gap bound) and economic (e.g., diminishing returns) properties (\S\ref{sec:algorithm}).
While in most of the paper we assume that the regulator knows exactly the preferences of each individual, we also study the case of imperfect information (\S\ref{sec:imperfect-information}).

Using data from the French census at the scale of a French department, we evaluate the \cotwo{} reduction achieved via the transportation mode incentive policy computed with our algorithm (\S~\ref{sec:case_study}).
The results show that our personalized incentives achieve the same \cotwo{} reduction as flat subsidies, but with a considerably smaller amount of incentives spent.
Our code is available as open source~\cite{JavaudinCode2022}.

We are aware that our framework is based on several idealized assumptions that makes its direct applicability difficult in practical situations, in particular for what concerns the assumption of being able to collect precise information about individual preferences.
In this sense, the path toward personalized-incentives is still a long way to go.
However, we argue that the theoretical findings of this paper, coupled with the continuous evolution of techniques for collecting societal big-data, while respecting privacy, provide important steps along this path.

%%%%%%%%%%%%%%%%%%%%%%%%%%%%%%%%%%%%%%%%%%%%%%%%%%%%%%%%%%%%%%%%%%%
%%%%%%%%%%%%%%% PROBLEM DEFINITION %%%%%%%%%%%%%%%%%%%%%%%%%%%%%%%%
%%%%%%%%%%%%%%%%%%%%%%%%%%%%%%%%%%%%%%%%%%%%%%%%%%%%%%%%%%%%%%%%%%%

%%%%%%%%%%%%%%%%%%%%%%%%%%%%%%%%%%%%%%%%%%%%%%%%%%%%%%%%%%%%%%%%
%%%%%%%%%%%%%%%%%%%%%%%%% FRAMEWORK AND INCENTIVE POLICY %%%%%%%
%%%%%%%%%%%%%%%%%%%%%%%%%%%%%%%%%%%%%%%%%%%%%%%%%%%%%%%%%%%%%%%%

\section{Framework and Incentive Policy}
\label{sec:policy}

%In this section, we introduce a simple discrete-choice framework and define our incentive policy to maximize social utility under a budget constraint. 
%We then express the optimization problem that we want to solve.
%Furthermore, we will show analytically that the social utility obtained from the algorithm is boundedly close to the theoretical maximum.}

\subsection{Model}
\label{sec:model-and-assumptions}

We consider a population $\IndSet \equiv \{1, \dots, \NumOfInd\}$ of $\NumOfInd$ individuals.
Each individual $\Ind\in\IndSet$ chooses an alternative $\Alt$ among an \emph{individual-specific choice-set} $\InitSetDef$.
For example, we can consider individuals choosing a mode of transportation to commute to their work.
In this case, the choice set could be $\InitSetDef=\{\text{car}, \text{walk}, \text{bike}, \text{public transit}\}$.
The choice set can be individual-specific so that if individual $\Ind$ owns a car but individual $\Ind'$ does not, we could have $\InitSetDef=\{\text{car}, \text{walk}, \text{bike}, \text{public transit}\}$ and $\InitSet{\Ind'}=\{\text{walk}, \text{bike}, \text{public transit}\}$.

Let $\TrDef>0$ be an incentive provided by the regulator to individual~$\Ind$, when she chooses alternative~$\Alt$. 
Since $\TrDef$ changes from an individual to another, such policy is \emph{personalized}.

A policy influences the individual choice since the proposed monetary transfers change her utilities.
The \emph{utility} $\IndUtAfterFuncDef$ of individual $\Ind$ when choosing alternative $\Alt\in\InitSetDef$ is given by
$
%\label{eq:utility_def}
    \IndUtAfterFuncDef = \IndUtDef + \TrDef,
$, 
where $\IndUtDef\in\mathbb{R}$ is the intrinsic utility (in the absence of policy).
Each individual $\Ind$ chooses an alternative $\AltEqDef$ which maximizes her utility
$
    %\label{eq:max-ind-ut}
    \AltEqDef \in \argmax_{\Alt} \IndUtAfterFuncDef.
$
Each alternative $\Alt$ of individual $\Ind$ is characterized by a \emph{social indicator} $\SocUtDef\in\mathbb{R}$ (e.g., the opposite of \cotwo{} emissions induced by the commutes).

We assume the regulator has perfect information: it knows exactly the intrinsic utilities $\{\IndUtDef\}\OfIndAltDef$ and social indicators $\{\SocUtDef\}\OfIndAltDef$ of all the alternatives, for all the individuals. We will relax this assumption in \S\ref{sec:imperfect-information}.

The alternative chosen by each individual $\Ind$ in the absence of policy (i.e., where $\TrDef=0$, $\forall\Ind, \Alt$) is called \emph{default alternative}, and denoted $\DefAltDef \defeq \argmax_{\Altt\in\argmax_\Alt\IndUtDef} \SocUt{\Ind}{\Altt}$.
In order to convince an individual $i$ to shift from its default alternative to any other alternative $\Alt$, it is necessary and sufficient for the regulator to provide an incentive $\WeightDef \defeq \IndUt{\Ind}{\DefAltDef} - \IndUtDef$, to compensation for the decrease of individual utility. 

\subsection{Maximum Social Welfare Problem}
\label{sec:maximum-social-welfare-problem}
The regulator has to decide, for each individual $\Ind$, which alternative to incentivize, which can be summarized by a binary decision variable $\DecDef$ that is equal to $1$ if the regulator wants to make individual $\Ind$ choose alternative $\Alt$, and $0$ otherwise. The incentive can be thus written as $\TrDef=\DecDef\cdot \WeightDef$.
The \emph{Maximum Social Welfare problem} is
\begin{align}
\label{eq:max-soc-ut}
            \max_{\{\DecDef\}\OfIndAltDef} & \sum_{\Ind\in\IndSet} \sum_{\Alt\in\InitSetDef} \SocUtDef \DecDef & \\
            \label{eq:budget}
            \text{s.t.} & \sum_{\Ind\in\IndSet} \sum_{\Alt\in\InitSetDef} \WeightDef \DecDef 
            \le \Budget & \\
                        \label{eq:only-one}
                        & \sum_{\Alt\in\InitSetDef} \DecDef = 1, & \Ind\in\IndSet \\
                        & \DecDef \in \{0,1\}, &\Ind\in\IndSet,\  \Alt\in\InitSetDef
\end{align}
The objective function~\eqref{eq:max-soc-ut} aims to maximize the social welfare, i.e., the sum of all social indicators, constraint~\eqref{eq:budget} indicates that the regulator cannot spend more than $\Budget$. Moreover, only one alternative per individual is incentivized~\eqref{eq:only-one}.

For any budget $\Budget$, we indicate with $\MaxSocUt{\Budget}$ the maximum of the social welfare, solution of problem~\eqref{eq:max-soc-ut}.

\subsection{Maximum Social Welfare Curve Problem}
\label{sec:maximum-social-welfare-curve-problem}

Suppose now that the regulator is endowed with a maximum budget $\Budget$ and that he can spend any budget in the interval $\TotIncSym\in[0,\Budget]$. To decide the exact amount of budget that is convenient to spend, it is useful to obtain the \emph{Maximum Social Welfare Curve} $\Curve^*_\Budget$, representing the maximum social welfare reachable, $\MaxSocUt{\TotIncSym}$, for any budget $\TotIncSym\in[0, \Budget]$, i.e.
$
    \Curve^*_\Budget = \left\{ \big(\TotIncSym, \MaxSocUt{\TotIncSym} \big) \:\middle|\: \TotIncSym \in [0, \Budget] \right\}
    \label{eq:max-soc-ut-curve}
$, 
which is clearly monotone non-decreasing (the larger the budget spent, the larger the social welfare reached).
Observe that, although a maximum budget $\Budget$ is available, the regulator may not want to indiscriminately spend it all, but may choose the actual budget to invest in incentives, based on several criteria.
For instance, the regulator may use the above curve to find the minimum budget needed to reach a certain social-welfare target (see the example of Fig.~\ref{fig:max_soc_welfare_curve}).

%%%%%%%%%%%%%%%%%%%%%%%%%%%%%%%%%%%%%%%%%%%%%%%%%%%%%%%%%%%%%%%%%%%%%%%%%%
%%%%%%%%%%%%%%%%%%%%%%% APPROX ALGO %%%%%%%%%%%%%%%%%%%%%%%%%%%%%%%%%%%%%%
%%%%%%%%%%%%%%%%%%%%%%%%%%%%%%%%%%%%%%%%%%%%%%%%%%%%%%%%%%%%%%%%%%%%%%%%%%

\section{Approximation Algorithm}
\label{sec:algorithm}

The MCKP problem, and thus the Maximum Social Welfare problem \eqref{eq:max-soc-ut}, is NP-hard~\cite{Kellerer2004} .
We provide in this section a polynomial time algorithm based on greedy algorithms from the Operations Research literature, which gives us solutions boundedly close to the optimum.

\subsection{Preliminary Steps}
\label{sec:preliminary-steps}
\begin{figure}[ht]
    \centering
    \includegraphics[width=0.3\textwidth]{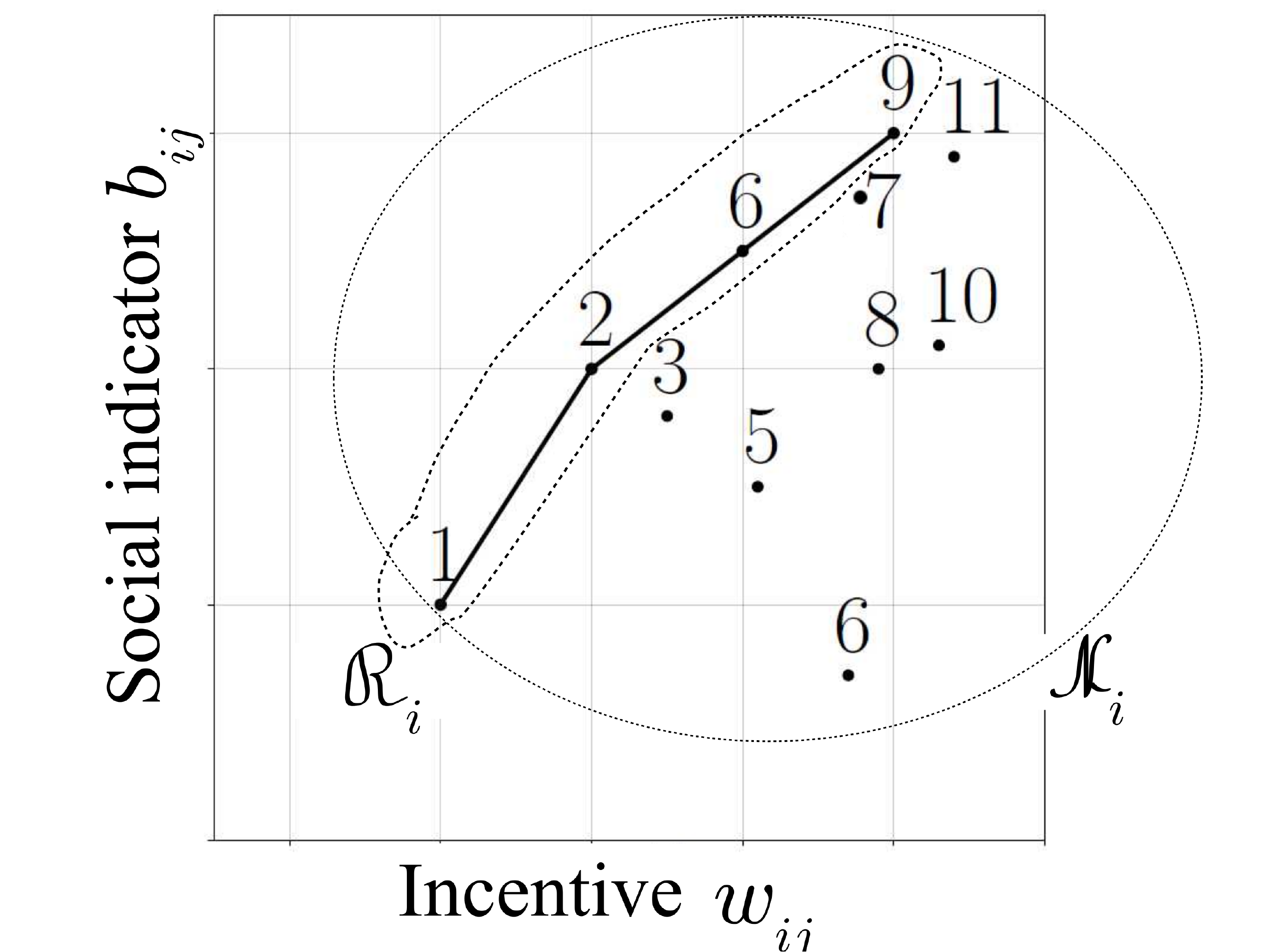}
    \caption{Alternative set $\InitSetDef$ of individual $\Ind$ and the subset $\NonDomSetDef$ of LP-extremes.}
    \label{fig:concavization}
\end{figure}

Before presenting the proposed algorithm, we need to ``clean'' the input of the problem, removing some irrelevant alternatives from the set $\InitSetDef$ of the alternatives of any individual $\Ind$~\cite[Section 11.2.1]{Kellerer2004}.
In broad terms, irrelevant alternatives are the ones that do not provide enough social indicator compared to the incentive amount needed to induce them.
The alternatives remaining after the cleaning are usually called \emph{LP-extremes} and we denote them with $\NonDomSetDef\subseteq\InitSetDef$.
Figure~\ref{fig:concavization} gives the intuition behind the process of constructing the set $\NonDomSetDef$, which is called \emph{concavization} \cite[Fig.1,2]{Zoltners1979}.
In the figure, alternative $3$ is irrelevant since $2$ provides a larger social indicator, while requiring less incentive.
Alternative $7$ is irrelevant since it requires to spend more incentive than $6$, for a negligible gain in the social indicator.
It is much more convenient to make a slightly bigger investment to induce alternative $9$, which provides a significant social indicator improvement with respect to $6$.

We follow the Operations Research literature in the slight abuse of notation of denoting with $\WeightDef$ the incentive to be provided to the $j$-th alternative in $\NonDomSetDef$.
With no loss of generality, we can assume the ordering 
$
    \label{eq:ordering}
    \Weight{\Ind}{1} < \Weight{\Ind}{2} < \dots < \Weight{\Ind}{\NumOfNonDomDef}
$. Obviously, the default alternative is the first alternative in the set $\NonDomSetDef$ and $\Weight{\Ind}{1}=0$.

\begin{definition}[Efficiency and incremental efficiency]
    \label{def:efficiency}
    We define the \emph{efficiency} of an alternative $\Alt$ of individual $\Ind$ as
    $
        \EffDef\defeq \frac{\SocUtDef - \SocUt{\Ind}{\DefAltDef} }{\WeightDef}
    $,
    i.e., the gain in social indicator that we can gain via a unit of incentive allocated to that alternative.
    
    We also define the incremental social indicator $\IncrSocUtDef$ and the incremental incentive $\IncrWeightDef$ required for each alternative $\Alt\in\NonDomSetDef$ as
    \begin{equation}
        \label{eq:incremental-weight-and-social}
        \begin{array}{rl}
            \IncrSocUtDef & \defeq \SocUtDef -\SocUt{\Ind}{\Alt-1} \\
            \IncrWeightDef  & \defeq \WeightDef -\Weight{\Ind}{\Alt-1}
        \end{array}, 
        \quad \Alt=2,\dots,\NumOfNonDomDef.
    \end{equation}
    The \emph{incremental efficiency} is then defined as
    $
        \label{eq:incremental_efficiency}
        \IncrEffDef \defeq \IncrSocUtDef / \IncrWeightDef.
    $
\end{definition}
The incremental efficiency $\IncrEffDef$ can be interpreted as the increase in social welfare for each monetary unit spent, when individual $\Ind$ shifts from alternative $\Alt-1$ to alternative $\Alt$.

\subsection{Greedy Algorithm}

Very efficient algorithms~\cite[Section~11.2.1]{Kellerer2004} are known to solve problem~\eqref{eq:max-soc-ut}, i.e., to approximate the maximum social welfare for a fixed single value of budget $\Budget$.
However, to apply them to the Maximum Social Welfare Curve problem, in which we want to find the maximum social welfare for a range of budget values $\TotIncSym\in[0,\Budget]$, instead of just one, we would have to run those algorithms from scratch for every single value of budget.
For this reason, we build our solutions upon a simpler greedy algorithm~\cite[Figure~11.2]{Kellerer2004}, which is less efficient to solve the Maximum Social Welfare problem (although still polynomial in time complexity), but easily extendable to also solve the Maximum Social Welfare Curve problem.
The other advantage deriving from such choice is that this greedy algorithm has interesting properties that increase its practical application.
%The regulator willing to apply such algorithms could choose, for instance, to apply an incentive policy up to the point where the marginal efficiency (marginal welfare divided by marginal cost) becomes less than a certain threshold, fixed based on political or societal considerations, as we do in Section~\ref{sec:case_study}.

The pseudocode of the algorithm is in Algorithm~\ref{alg:greedy-max-soc-ut-curve}.
The notation $\AltPairDef$ stands for ``$\Alt$-th alternative of individual $\Ind$''.
First, the algorithm finds all the LP-extremes alternatives and sort them by order of decreasing incremental efficiency.
Then, at each iteration, the next pair $\AltPair{\Ind'}{\Alt'}$ with the highest incremental efficiency is picked (line~\ref{ln:incr-eff}).
The alternative induced to $\Ind'$ is set to $\Alt'$ (line~\ref{ln:choice-update}) and the budget is reduced by the amount of the incremental weight~\eqref{eq:incr-weight}.
An additional piece of the approximation of the social welfare curve is computed~\eqref{eq:AlgSocUt}.
%The algorithm stops when the budget is depleted.
%From the allocation $\{\Dec{\Ind}{\Alt}\}\OfIndAltDef$ returned by the algorithm, we can deduce a state $(\Alt_1, \dots, \Alt_\NumOfInd)$, where $\Dec{\Ind}{\Alt_\Ind} = 1$, for any individual $\Ind$.
%Then, for any individual $\Ind$, the regulator proposes an incentive $\Inc{\Ind}{\Alt_\Ind}$ for alternative $\Alt_\Ind$ and no incentive for any alternative $\Alt\neq\Alt_\Ind$.
%Some individuals might receive no incentive if $\Dec{\Ind}{\DefAltDef}=1$ (the allocation returned by the algorithm is such that they choose their default alternative).
The algorithm stops when the maximum budget $\Budget$ is depleted.

\begin{algorithm}[ht!]
\caption{Greedy algorithm for the Maximum Social Welfare and Maximum Social Welfare Curve problems.}
\label{alg:greedy-max-soc-ut-curve}
\footnotesize
\SetKwInOut{Input}{Input}\SetKwInOut{Output}{Output}
\Input{Social indicators $\{\SocUtDef\}\OfIndAltDef$, intrinsic utilities $\{\IndUtDef\}\OfIndAltDef$, budget $\Budget$ }
Iteration index $\IterSym:=0$\\
$\TotInc\OfIterDef:=0$, Total incentive allocated so far.\\
$\TotSocUt\OfIterDef:=0$, Social welfare obtained in the current allocation.\\
Compute the ordered set $\NonDomSetDef$ of LP-extremes of each individual $\Ind$.\\
Sort all the alternatives $\AltPairDef$ according to decreasing incremental efficiency $\IncrEffDef$ and put them in a set $\TotNonDomSet$.\\
Initialize the alternatives chosen by the individuals $\{\DecDef\}\OfIndAltDef$ as follows
\begin{equation*}
    \left\{
    \begin{array}{ll}
        \Dec{\Ind}{1} = 1, & \text{(default alternative)} \\
        \DecDef = 0, & \text{for any alternative }\Alt>1
    \end{array}
    \right.
\end{equation*}
\\
\While{$\TotNonDomSet\neq\emptyset$ and $\TotInc\OfIterDef\le \Budget$}
{
	Take $\AltPair{\Ind'}{\Alt'}$, the next alternative with the highest incremental efficiency $\IncrEff{\Ind'}{\Alt'}$ from $\TotNonDomSet$. \label{ln:incr-eff}\\ 
	Add $\AltPair{\Ind'}{\Alt'}$ to the solution, i.e.:
	\begin{align}
	\TotNonDomSet & := \TotNonDomSet\setminus\{\AltPair{\Ind'}{\Alt'}\}, & \nonumber
	\\
	\TotInc\OfIter{\IterSym+1} 
	& := \TotInc\OfIterDef + \IncrWeight{\Ind'}{\Alt'}
        \label{eq:incr-weight}
	\\
	\IncrEffSym\OfIterDef
	& := \IncrEff{\Ind'}{\Alt'}
	\label{eq:incr-eff-iter}
	\\
	\AlgSocUtDef
	& :=\TotSocUt\OfIterDef, 
	& \forall \TotIncSym\in [\TotInc\OfIterDef,\TotInc\OfIter{\IterSym+1})
	\label{eq:AlgSocUt}
	\\
	\TotSocUt\OfIter{\IterSym+1} 
	&:= \TotSocUt\OfIterDef + \IncrSocUt{\Ind'}{\Alt'} 
	& \nonumber \\
	\IterSym 
	& := \IterSym +1 \nonumber
	&
	\end{align}
	\label{ln:iteration-wise}
	\\
	Update the selected alternative for individual $\Ind'$, i.e.,
	\label{ln:choice-update}
\begin{equation*}
    \left\{
        \begin{array}{rll}
            \Dec{\Ind'}{\Alt'} & = 1, &  \\
            \Dec{\Ind'}{\Alt} & = 0, & \text{ for any other alternative }\Alt\neq \Alt'
        \end{array}
    \right.
\end{equation*}
}
%
%\rev{aa}{$\SocWelfareSym(\Budget) := \TotSocUt\OfIter{\IterSym+1}, \forall \Budget\in[\TotInc\OfIter{\IterSym+1},+\infty[$}{} \\
%
\Output{
Curve $\Curve_\Budget=\left\{(\TotIncSym,\AlgSocUtDef)\:|\:\TotIncSym\in[0,\Budget] \right\}$\\
Chosen alternatives $\{\DecDef\}\OfIndAltDef$\\
Incentive policy $\IncPolicy = \{\IncDef\}\OfIndAltDef$, where $\IncDef=\DecDef\cdot\WeightDef$\\
Split item $\AltPair{\SplitInd}{\SplitAlt}:=\AltPair{\Ind'}{\Alt'}$\\
Incremental efficiency of the split item $\IncrEff{\SplitInd}{\SplitAlt}$\\
Budget actually used $\BudgetUsed:=\TotInc\OfIter{\IterSym-1}$
}
\end{algorithm}

Observe that the curve $\Curve_\Budget$ given as output by the algorithm is an approximation of the solution $\Curve^*_\Budget$ of the Maximum Social Welfare Curve Problem (Section~\ref{sec:maximum-social-welfare-curve-problem}).
Moreover, given any maximum budget $\Budget$, the algorithm returns an approximation $\AlgSocUt{\Budget}$ to the solution $\MaxSocUtDef$ of the Maximum Social Welfare Problem~\eqref{eq:max-soc-ut}.
Note that, in order to achieve $\AlgSocUt{\Budget}$, the policy issued by the algorithm does not spend the entire maximum budget $\Budget$, but only $\BudgetUsed\le\Budget$.

The algorithm also gives as output the incremental efficiency of the ``split item'', denoted with $\IncrEff{\SplitInd}{\SplitAlt}$, useful to compute the optimality gap of the algorithm (Theorem~\ref{thm:bound} below).
The name \emph{split item}, which we borrow from~\cite{Kellerer2004}, reminds that, when we allocate budget $\Budget$, we add to the solution all the LP-extreme alternatives, in decreasing order of incremental efficiency, up to the ``split alternative'' $\AltPair{\SplitInd}{\SplitAlt}$, as $\AltPair{\SplitInd}{\SplitAlt}$ actually splits the set $\TotNonDomSet$ of all LP-extremes in two subsets: the first containing the alternatives to include in our solution, while we do not include the LP-extremes from the second subset.

The following statements guarantee that the result from the algorithm is boundedly close to the optimum. We omit the proofs for lack of space, which can be obtained by adapting~\cite[Ch.11]{Kellerer2004}.

\begin{theorem}[Upper bound]
    \label{thm:bound}
    Let us run Alg.~\ref{alg:greedy-max-soc-ut-curve} with budget $\Budget$, and let $\BudgetUsed$ be the budget actually used and $\IncrEff{\SplitInd}{\SplitAlt}$ be the incremental efficiency of the split item.
    The social welfare $\AlgSocUt{\Budget}$ we obtain is boundedly close to the social welfare $\MaxSocUtDef$ of any optimal personalized-incentive policy:
    \begin{align}
        \label{eq:bound}
        \MaxSocUtDef - \AlgSocUt{\Budget}
        \le \IncrEff{\SplitInd}{\SplitAlt} \cdot (\Budget-\BudgetUsed).
    \end{align}
\end{theorem}

\begin{corollary}
    \label{prop:curve_bound}
    The curve $\Curve_\Budget$ obtained via Algorithm~\ref{alg:greedy-max-soc-ut-curve} is boundedly close to the Maximum Social Welfare Curve $\Curve^*_\Budget$ from~\eqref{eq:max-soc-ut-curve} and the gap is given by Theorem~\ref{thm:bound}.
\end{corollary}

\begin{figure}[ht]
    \centering
    \includegraphics[width=0.5\textwidth]{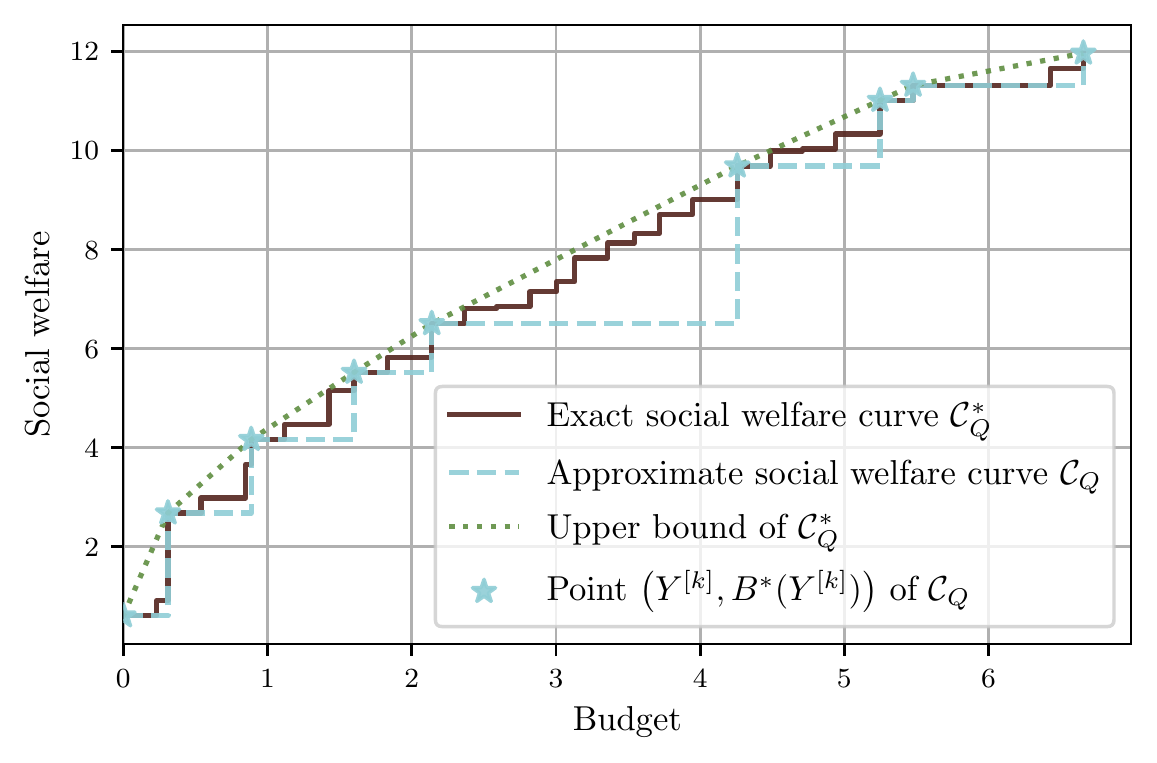}
    \caption{Distance between the social welfare curve $\Curve_\Budget$ computed by Algorithm~\ref{alg:greedy-max-soc-ut-curve}, the maximum social welfare curve $\Curve^*_\Budget$ (Section~\ref{sec:maximum-social-welfare-curve-problem}) and the upper bound of Theorem~\ref{thm:bound}. 
    The stars represent the incentive spent $\TotInc\OfIterDef$ and social welfare $\TotInc\OfIterDef = \MaxSocUt{\TotInc\OfIterDef}$ at each iteration $\IterSym=1,\dots,8$ of the algorithm (line~\ref{ln:iteration-wise}).}%
    \label{fig:distance_optimum}
\end{figure}

Fig.~\ref{fig:distance_optimum} illustrates the property above in a small example.

\begin{proposition}
\label{prop:complexity}
The computational complexity of Alg.~\ref{alg:greedy-max-soc-ut-curve} is $O(\sum_{\Ind=1}^\NumOfInd \NumOfAltDef \cdot \log \NumOfConvHullDef + \TotNumOfConvHull \cdot \log m)$, where $\NumOfInd$ is the number of individuals, $\NumOfAltDef$ is the number of alternatives of individual $\Ind$, $\NumOfConvHullDef$ is the number of LP-extremes of individual $\Ind$ and $\TotNumOfConvHull\defeq\sum_{\Ind=1}^\NumOfInd \NumOfConvHullDef$.
\end{proposition}
%Note that the component $\sum_{\Ind=1}^\NumOfInd \NumOfAltDef \cdot \log \NumOfConvHullDef$ is for computing the LP-extremes, using the method of \cite{Kirkpatrick1986}.
Note that, since the alternatives of each individual are independent of the others, the sets $\NonDomSetDef$ can be computed in parallel, thus reducing even further the computation time.
%Therefore, the computation performance shown in the numerical results, in which we do not use any parallelization, should be considered as a pessimistic bound.

Despite our algorithm being computationally efficient, there might be cases in which it is desirable to stop it prematurely, without waiting for it to completely terminate.
This can be the case when a personalized-incentive policy must be computed on-the-fly, within tight time-constraints.
The following properties ensure that our algorithm is suitable to this situation, which eases its practical adoption.

\begin{remark}[Anytime algorithm]
\label{rem:anytime}
Alg.~\ref{alg:greedy-max-soc-ut-curve} is \emph{anytime}: if we stop it prematurely at any iteration $\IterSym$, we get a valid solution for the Maximum Social Welfare and the Maximum Social Welfare Curve problems, with budget $\Budget'=\TotInc\OfIterDef$.
\end{remark}

%The property of Prop. \ref{prop:incr_optimality} ensures that all the solutions that the algorithm incrementally builds are optimal, with respect to the expenses of that allocation.
\begin{remark}[Incremental use]
Another desirable property of Algorithm~\ref{alg:greedy-max-soc-ut-curve} is that we can build on a previously computed incentive allocation whenever new available budget becomes available, instead of recomputing the entire allocation from scratch.
To explain this, let us suppose that we have a certain budget $\Budget$ and the algorithm returns the allocation $\{\DecDef\}\OfIndAltDef$, spending the corresponding incentive amount $\BudgetUsed$.
Suppose now that the available budget increases to $\Budget'>\Budget$. In this case, in order to exploit the new additional budget, we can simply resume the algorithm from its last iteration and continue up to the furthest iteration such that $\TotInc\OfIter{\IterSym+1}\le\Budget'$.
This is, per-se, a computational advantage with respect to algorithms that need to run from scratch every time new resources (budget) are available.
\end{remark}

%%%%%%%%%%%%%%%%%%%%%%%%%%%%%%%%%%%%%%%%%%%%%%%%%%%%%%%%%%%%%%%%%%%%%%%%%%
%%%%%%%%%%%%%%%%%%%%%%%%%% IMPERFECT INFORMATION %%%%%%%%%%%%%%%%%%%%%%%%%
%%%%%%%%%%%%%%%%%%%%%%%%%%%%%%%%%%%%%%%%%%%%%%%%%%%%%%%%%%%%%%%%%%%%%%%%%%

\section{Imperfect Information}
\label{sec:imperfect-information}

The assumption that the regulator knows perfectly the utility of the individuals may seem restrictive.
In this section, we show that the algorithm is still relevant when the utility is imperfectly known.
From discrete-choice theory \cite{Anderson1992}, we assume that intrinsic utility of alternative $\Alt$ of individual $\Ind$ is $\IndUtDef = \IndUtDetDef + \RandomDef$, where $\IndUtDetDef$ is deterministic  and $\RandomDef$ randomly Gumbel-distributed. Their specific parameters are specified in our code~\cite{JavaudinCode2022}.

We assume that the regulator knows the deterministic part $\IndUtDetDef$ of the utility but not the random part $\RandomDef$.

Under this assumption, the regulator does not know the minimum incentive amount needed to induce individual $\Ind$ to shift from her default alternative $\DefAltDef$ to another alternative $\Alt$.
A heuristic solution would be to set the incentive amount equal to the expectation of the utility difference between the two alternatives, given that $\DefAltDef$ is the default alternative chosen when there is no incentive.
%Observe that, by construction (see equation~\eqref{eq:default_alt}), $\IndUt{\Ind}{\DefAltDef} > \IndUtDef$.
In this case, the incentives $\{\Inc{\Ind}{\Alt}\}_{\Alt\in\InitSetDef}$ proposed by the regulator to individual~$\Ind$, to convince her to shift to alternative~$\Alt$, are such that $\Inc{\Ind}{\Altt}=0$, for any $\Altt\neq\Alt$, and
\begin{align}
    \label{eq:incentive_amount_imperfect_0}
    \IncDef = & \ExpSym(\IndUt{\Ind}{\DefAltDef} - \IndUtDef |\IndUt{\Ind}{\DefAltDef} > \IndUtDef) =
    \\
    &\IncDetDef + \ExpSym(\Random{\Ind}{\DefAltDef} - \RandomDef | \Random{\Ind}{\DefAltDef} - \RandomDef > -\IncDetDef),
\end{align}
where $\IncDetDef = \IndUtDet{\Ind}{\DefAltDef}-\IndUtDetDef$ is the difference in the deterministic part of the utility, known to the regulator.

Given an individual $\Ind$ and an alternative $\Alt\in\InitSetDef$, if the regulator proposes the incentive $\IncDef$, as defined by~\eqref{eq:incentive_amount_imperfect_0}, then individual $\Ind$ has a positive probability to refuse the incentive.
Hence, the expenses of the regulator may be smaller than the total incentive amount proposed.

Algorithm~\ref{alg:greedy-max-soc-ut-curve} can be used to compute a personalized-incentive policy under imperfect information, by defining new weights
\begin{equation*}
    %\WeightDef = \mu\frac{ 1 + e^{\IncDetDef/\mu} }{ e^{\IncDetDef/\mu} } \ln\left( 1 + e^{\IncDetDef/\mu} \right), \quad \forall\Ind, \Alt.
    \WeightDef = \ExpSym(\IndUt{\Ind}{\DefAltDef} - \IndUtDef |\IndUt{\Ind}{\DefAltDef} > \IndUtDef).
\end{equation*}

At each iteration of the algorithm, the regulator proposes the incentive $\Weight{\Ind'}{\Alt'}$ to individual $\Ind'$ for alternative $\Alt'$, where $[\Ind', \Alt']$ is the pair of individual and alternative selected by the algorithm.
The regulator observes the response of the individual to the incentive.
If the individual accepts the incentive, it decreases the budget by the incentive amount.
The regulator keeps proposing incentives one by one until his budget is depleted.

Note that, if an individual $\Ind$ accepts an incentive $\IncDef$ for alternative $\Alt\in\InitSetDef$, the regulator can still propose  her, later, an incentive $\Inc{\Ind}{\Altt}$ for another alternative $\Altt\in\InitSetDef$.
If the individual refuses the second incentive $\Inc{\Ind}{\Altt}$, she still receives the first incentive $\IncDef$. 

In Section~\ref{sec:modal-choice-imperfect-information}, we apply the policy presented above to our case study and compare it to the case with perfect information, assuming that random terms are Gumbel-distributed.
The following proposition gives the exact expression of the incentives~\eqref{eq:incentive_amount_imperfect_0}, in case of Gumbel-distributed random terms.
\begin{proposition}
\label{prop:gumbel}
Let us assume that the random terms are i.i.d.~and follow a Gumbel distribution with scale parameter $\mu$ (i.e., $\RandomDef/\mu$ follows a standard Gumbel distribution).
%It is well known \cite{Nadarajah2005} that the difference of two i.i.d. Gumbel-distributed random variables is a standard logistic-distributed random variable, whose probability density function is $f(x)=\frac{e^x}{(e^x+1)^2}$.
Then, the incentive amount from~\eqref{eq:incentive_amount_imperfect_0} can be written as
\begin{equation*}
    %\label{eq:incentive_amount_imperfect}
    \IncDef = \mu \frac{ 1 + e^{\IncDetDef/\mu} }{ e^{\IncDetDef/\mu} } \ln\left( 1 + e^{\IncDetDef/\mu} \right) \geq 0.
\end{equation*}
\end{proposition}

 \iffalse

 %==========
 % Probability of acceptance
 %==========

 The probability that individual $\Ind$ accepts the incentive $\IncDef$ proposed to her is
 \begin{equation*}
     \AccProbDef = \ProbSym(\IndUtDef + \IncDef \geq \IndUt{\Ind}{\DefAltDef} | \IndUt{\Ind}{\DefAltDef} > \IndUtDef).
 \end{equation*}
 Using equation \eqref{eq:random-ind-ut}, we get
 \begin{equation*}
     \AccProbDef = \ProbSym(\Random{\Ind}{\DefAltDef} - \RandomDef \leq \IncDef - \IncDetDef | \Random{\Ind}{\DefAltDef} - \RandomDef > -\IncDetDef).
 \end{equation*}
 With $\xi=\Random{\Ind}{\DefAltDef} - \RandomDef$,
 \begin{equation*}
     \AccProbDef = 1 - \ProbSym(\xi > \IncDef - \IncDetDef | \xi > -\IncDetDef) = 1 - \frac{1-F(\IncDef - \IncDetDef)}{1 - F(-\IncDetDef)}.
 \end{equation*}
 Then, using the cumulative distribution function of $\xi$,
 \begin{equation*}
     \AccProbDef = 1 - \frac{1+e^{-\IncDetDef/\mu}}{1+e^{(\IncDef-\IncDetDef)/\mu}} = \frac{e^{(\IncDef-\IncDetDef)/\mu}-e^{-\IncDetDef/\mu}}{1+e^{(\IncDef-\IncDetDef)/\mu}}.
 \end{equation*}
 Rearranging the terms yields
 \begin{equation}
     \label{eq:acceptance_prob}
     \AccProbDef = \frac{1-e^{-\IncDef/\mu}}{1+e^{(\IncDetDef-\IncDef)/\mu}}.
 \end{equation}

 \fi

\section{Application to Mode Choice}\label{sec:case_study}

We consider a regulator willing to employ a limited monetary budget in order to promote eco-friendly modes of transportation in order to reduce \cotwo{} emissions.

Observed variables include city- or district-level home and work location, main mode of transportation used for commuting, and some socio-demographic variables.
The modes of transportation are: car, public transit, walking, cycling and motorcycle. To estimate the utilities $\IndUtDef$ of each mode of transportation perceived by each individual, we resort to multinomial logit modeling on census data on the Rhône department (\num{222000} households) from the French statistics institute INSEE in 2015-19. To estimate the social indicators $\SocUtDef$, we use data from the French Environmental Agency~\cite{ADEMEGreenHouse}. Details about the datasets and the estimation procedure can be found in our repository~\cite{JavaudinCode2022}.

We implicitly assume that the utility and the social benefit of an individual when commuting by car or public transit does not depend on how many other individuals commute by car or by public transit.
This approximation is legitimate if the number of modal shifts induced by the policy is low, so that their impact on congestion and occupation is negligible. We checked a posteriori that this latter assumption is verified in our case (less than \SI{1.60}{\%} of individuals shifted mode).

\subsection{Calculation of the Personalized-Incentive Policy}
\label{sec:calculation-of-the-incentive-policy}
\begin{figure}
    \centering
    \includegraphics[width=\columnwidth]{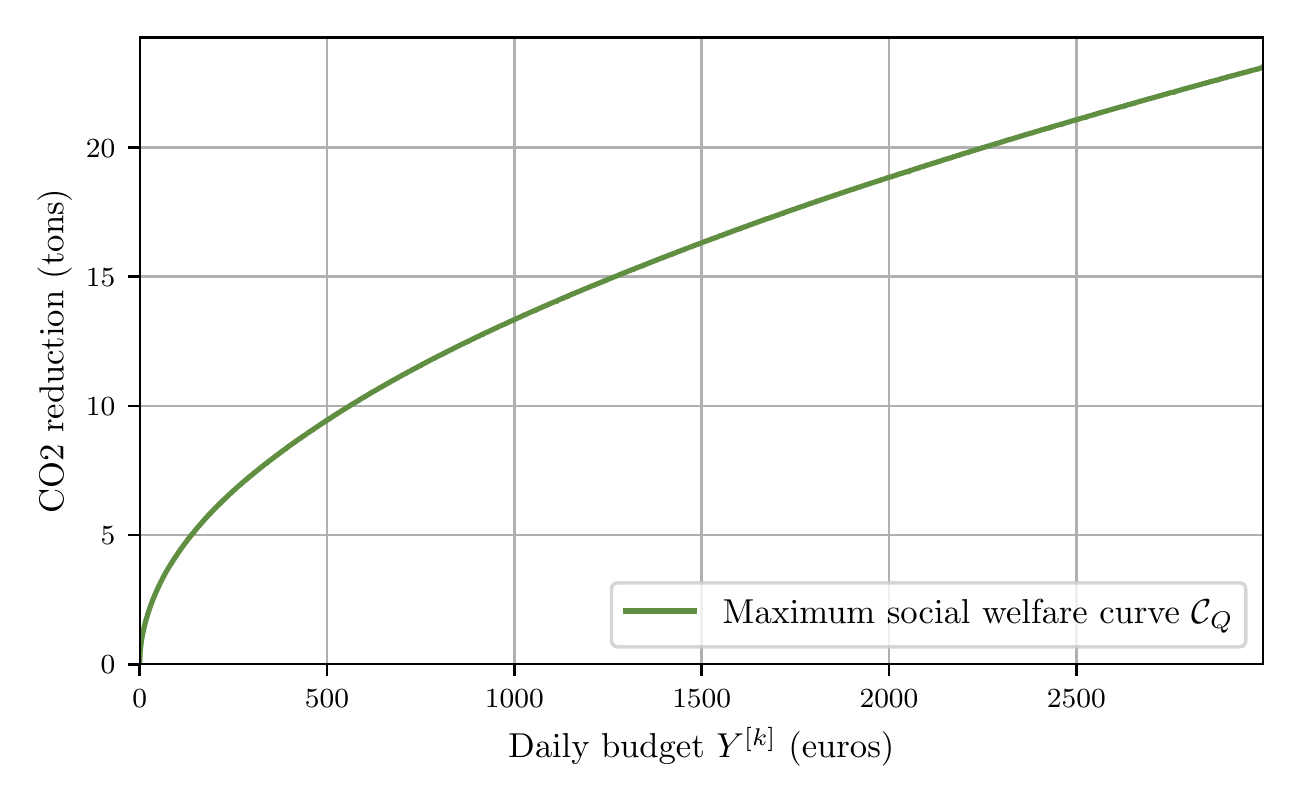}
    \caption{Maximum social welfare curve, up to a daily budget of 3000~€. \textbf{Note}: The social welfare corresponds to the reduction in \cotwo{} emissions due to the personalized-incentive policy.}%
    \label{fig:max_soc_welfare_curve}
\end{figure}

%\todo[inline]{aa: one thing over which the reviewer could attack us is ``you did not consider electric vehicles''}
We have about $2\cdot 10^5$ individuals and $10^6$ alternatives. The regulator proposes, each day, incentives to the individuals before their home-work trip. The social indicator of an alternative is the reduction in \cotwo{} emissions for the trip \emph{back and forth}, with respect to the default alternative.
The budget is the daily amount available to the regulator.
The social welfare curve given by Alg.~\ref{alg:greedy-max-soc-ut-curve} when daily budget is \num{3000}~€ is in Fig.~\ref{fig:max_soc_welfare_curve}.

We then set the budget of the regulator to $\Budget=\num{1800}$~€.
Running Algorithm~\ref{alg:greedy-max-soc-ut-curve} with this budget required about \num{3500} iterations and took about 6 seconds (with Python, on a computer with an Intel i5-8350U 1.7GHz and 24GB of memory).
The algorithm allocates practically all the budget (\num{1798.59}~€), inducing modal shift of~\num{1.57}\% of individuals and \cotwo{} reduction by \num{18} tons per day (\SI{3.00}{\%} of total \cotwo{} emissions).
Thus, this policy would cost on average \num{100.61}~€ for each ton of \cotwo{} prevented, which is a reasonable carbon price~\cite{quinet2009valeur}.
%\todo[inline]{aa: to justify the polynomial time algorithm, we had discussed to compare its running time with that of a solver. Lucas, do you have such a comparison?\\
%lj: It runs in 500 seconds with SCIP solver (so roughly 50-100 times longer).\\
%aa: mmmh, 500 seconds would a sufficient time for the regulator to compute the incentives for the entire day! Therefore, in our studied case, the existence of our algorithm is not really justified. Such an algorithm would be more relevant at a bigger scale, when the solution from the solver would need about 10 hours or in an application with a much tighter time-constraints, for instance when decisions must be taken with a delay of 10 seconds.\\ Therefore, it is better to leave the information about the solver running time out of the paper and, if the reviewer ask, we will add bigger scenarios, multiplying by 2, 4, 8. 16 the population and plotting the computation time of our algorithm vs. the solver. But let's keep this for later.}

\begin{figure}
    \centering
    \includegraphics[width=\columnwidth]{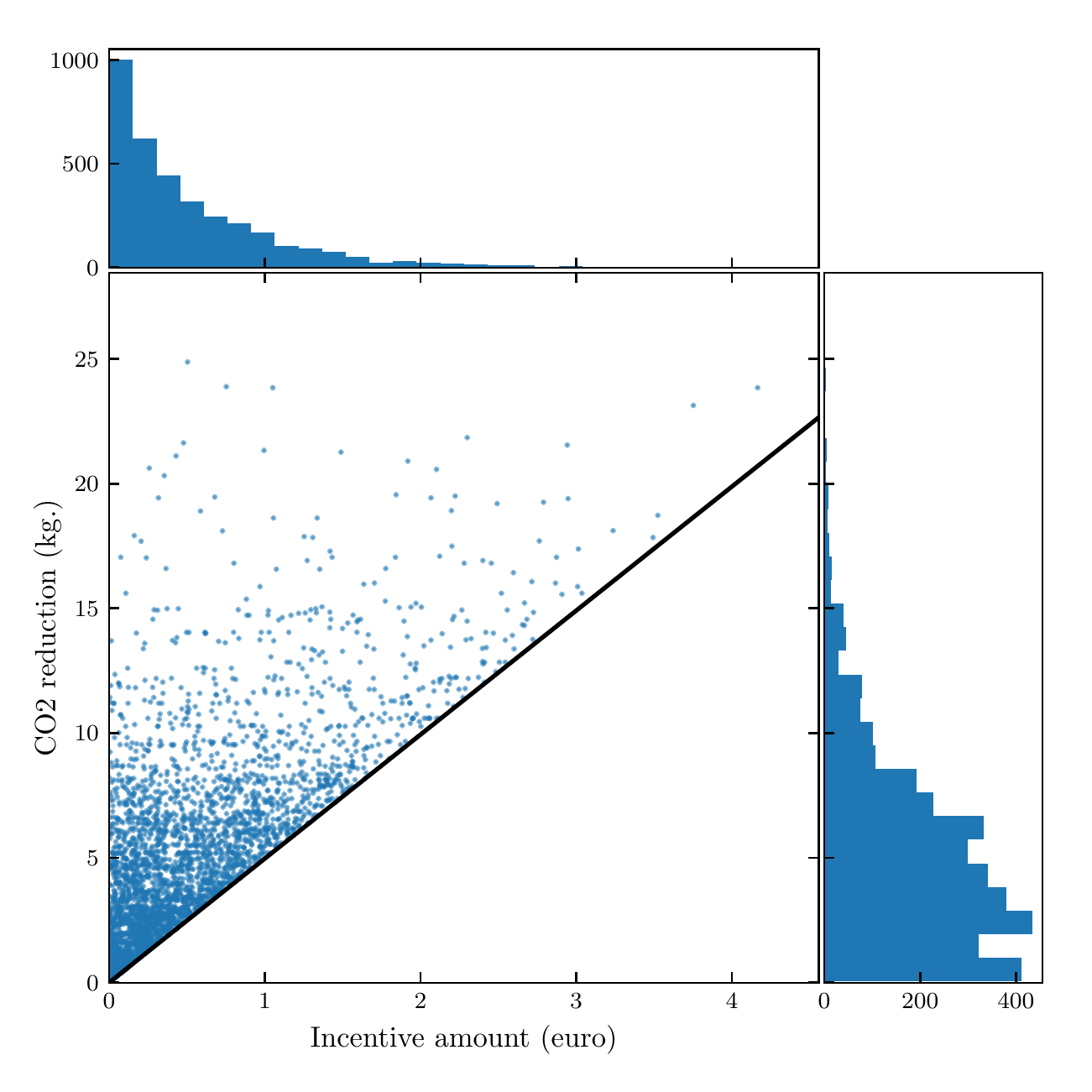}
    \caption{Distribution of incentive amount and \cotwo{} reduction for the incentives given in one day with budget $\Budget=1800$~€.\\
    The slope of the black line represents the incremental efficiency of the split item returned by the algorithm, $\IncrEff{\SplitInd}{\SplitAlt}=\num{5}$ tons of \cotwo{} / euro.
    Note that all points are above the line because their incremental efficiency is larger.
    The histogram above represents the distribution of the incentive amounts.
    The histogram on the right represents the distribution of the \cotwo{} reduction for the incentives.}%
    \label{fig:jumps_scatter}
\end{figure}

Despite the small incentives, the reduction in \cotwo{} emissions is considerable.
Indeed, among the individuals who received incentives, the average amount of incentives is \num{0.52}~€ per individual, for an average daily reduction in \cotwo{} emissions of \num{5} kilograms.
Recall that alternatives providing a large reduction in \cotwo{}, while requiring small incentive, have a high efficiency.
Hence, the algorithm selects first shifts achievable with a small incentive, i.e., where the individual is almost indifferent between the two alternatives, which however have a large difference in \cotwo{}.
Fig.\ref{fig:jumps_scatter} shows the distribution of the incentive amount and the \cotwo{} reduction for the incentivized individuals.
For most incentives, the amount proposed to individuals is below \num{1} euro (larger incentives are rarely efficient).

%On average, the reduction in \cotwo{} emissions from the incentives is \num{1.89} kilograms, while the average individual in the population emits only \num{1.28} kilograms per trip and the median is \num{0.65} kilogram.
%The share of car commutes decrease (from \SI{65}{\%} to \SI{45}{\%}) as the share of commutes by public transit (from \SI{18}{\%} to \SI{20}{\%}), by foot (from \SI{10}{\%} to \SI{13}{\%}) and by bike (from \SI{7}{\%} to \SI{22}{\%}) increase.
%Interestingly, the mode of transportation with the highest increase in use is the bike, which suggests most car drivers prefer to switch to the bike instead of walking.
%\todo[inline]{aa: I think this is not realistic. Nobody would expect that all of the sudden everyones gets on a bike. We limited the car possibility only to individuals owning a car. Maybe we should also limit the use of bike only to the individuals using the bike. Are we considering velib or similar? I am pretty sure that among the individuals switching to bikes, there are poor guys doing 30 Km by bike. Is there any way in the model to avoid this? Are we sure it does not happen. To control this does not happen, we should have a scatterplot like in Fig.~\ref{fig:transition}.
 %}
 %\begin{figure}
 %\includegraphics[width=0.7\textwidth]{graphs/andrea/transition}
 %\caption{In this plot, each point represents an individual and its transition, i.e., the mode she was choosing when 0 incentive budget is invested (her default choice) and the mode chosen when the incentive policy is implemented.}
 %\label{fig:transition}
 %\end{figure}

\begin{figure}
    \centering
    \includegraphics[width=\columnwidth]{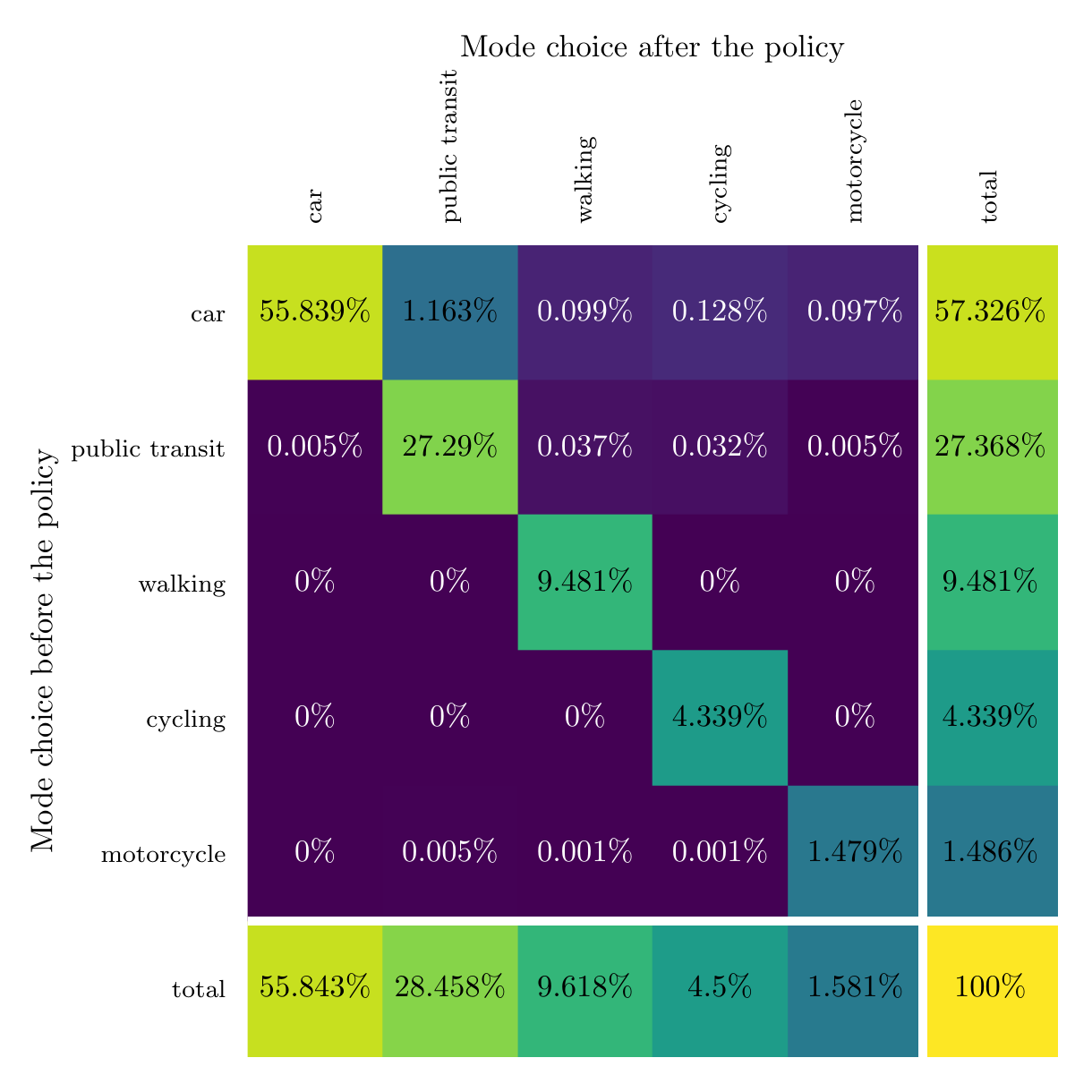}
    \caption{Evolution of mode share before and after the policy. \SI{1.163}{\%} of individuals were given incentives to shift from car to public transit, \SI{27.29}{\%} of individuals commuted by public transit before the policy and were not induced to shift.}%
    \label{fig:switches_heatmap}
\end{figure}

Fig.\ref{fig:switches_heatmap} compares mode share before and after the policy.
Most individuals who received incentives are individuals who commuted by car and were induced to commute by public transit (\num{1.2}\% of all individuals, \num{74}\% of individuals who received incentives).
The share of individuals commuting by car decreased by 2.4\%, while public transit ridership increased by \num{4}\%.

%The incremental efficiency of the last iteration is \num{0.24} kilogram of \cotwo{} per euro.
%Hence, from Proposition \ref{prop:taxation}, the same results could be obtained with a proportional tax on the \cotwo{} emissions emitted by the individuals during their trips.
%The tax level should be set to $\Tax = 1/0.24 \approx 4.14$ euros per kilogram of \cotwo{} emitted, which is about 1.05 euros per kilometer by car.
%Such taxation would bring a revenue of \num{8390929.09} euros for the regulator but the sum of the individual utilities (expressed in euros) would decrease by \num{8390929.09} + \num{9999992.65} = \num{18390921.74} euros.
%%\todo[inline]{aa: What do you mean by ``the same results''? They seem not to be the same, as you say that the individuals would loose more utility. Please, specify better.}

We now compute a bound of the optimality gap, i.e., the maximum additional \cotwo{} savings we would achieve if we could use a theoretical optimal policy instead of resorting to Algorithm~\ref{alg:greedy-max-soc-ut-curve} (Theor.~\ref{thm:bound}).
Since the incremental efficiency of the split item returned by the algorithm is $\IncrEff{\SplitInd}{\SplitAlt}\simeq 5$ kilograms of \cotwo{} per euro and the unused budget is $\Budget-\BudgetUsed=1.41$~€, an optimal policy would reduce of just $5\cdot 1.41 \simeq 7$ kilograms more than Algorithm~\ref{alg:greedy-max-soc-ut-curve}, which is negligible compared to the total \cotwo{} emissions reduction of \num{18} tons provided overall.

\subsection{Imperfect Information}
\label{sec:modal-choice-imperfect-information}

We show in this section the performance of our allocation policy when the regulator has imperfect information about individual utilities.
In this case, the allocation policy is computed as in Section~\ref{sec:imperfect-information}.
Using the values of the random variables $\RandomDef$ drawn previously, we can check whether individuals accept the incentives proposed to them.
The policy stops when the daily budget of \num{1800}~€ is depleted.

\begin{table}
    \centering
    \caption{Performance of the personalized-incentive policy for one day, with perfect and imperfect information.}%
    \label{tab:perfect-vs-imperfect}
    \begin{tabular}{lcc}
        \toprule
        & Perfect information & Imperfect information \\
        \midrule
        Budget spent & \num{1798.59}~€ & \num{1797.03}~€ \\
        Incentives proposed & \num{3486} & \num{419}\\
        Incentives accepted & \num{3486} & \num{247}\\
        Acceptance rate & \SI{100}{\%} & \SI{59}{\%} \\
        \cotwo{} reduction & \num{17.9} tons & \num{3.8} tons \\
        \bottomrule
    \end{tabular}
\end{table}

Table~\ref{tab:perfect-vs-imperfect} compares the performance of our personalized-incentive policy under perfect and imperfect information.
As expected, imperfect information decreases the efficacy of the policy.
Since the regulator does not exactly know the individual utilities, it may propose insufficient incentives, which are rejected by individuals (it happens \SI{41}{\%} of the times).
This results in a smaller reduction of \cotwo{} (\SI{21}{\%} compared with the perfect information case).
Note that less individuals are involved in the incentive program (only \SI{12}{\%} compared to the perfect information case) because incentive given to single individuals are on average larger, and thus the budget is depleted more quickly.

These results could be improved by learning from the responses of individual $\Ind$ to the incentives proposed earlier in order to compute the incentives that will be proposed to her for other alternatives.
For example, if the regulator observes that individual $\Ind$ refused the incentive to shift from car to walking, he learns information on the random term of the utility for car of individual $\Ind$.

Also, if it is not possible to propose incentives to individual $\Ind$ for different alternatives consecutively, the regulator could propose incentives for multiple alternatives simultaneously.

These extensions cannot be carried out with Algorithm~\ref{alg:greedy-max-soc-ut-curve}.
Future work could study the optimal personalized-incentive policy under imperfect information.

\section{Conclusion}\label{sec:conclusion}

% imperfect information
% multi-criteria problem
% multi-constraint problem
% strategic behavior / unfairness (pollutants get more incentives)

This paper explores a computationally efficient method for the regulator to determine the optimal incentives to be provided to each individual to alter their choices in order to maximize social welfare. Such a method requires to know the preferences of individuals, which could be possible, thanks to the wealth of data available for user nowadays, in compliance to privacy.
We will more systematically study in our future work the imperfect information case, which we have here tackled empirically. Moreover, we will consider the possible congestion induced by the distribution of incentives, possibly via an iterative procedure alternating the incentive algorithm and the computation of the current level of congestion.

%%%%%%%%%%%%%%%%%%%%%%%%%%%%%%%%%%%%%%%%%%%%%%%%%%%%%%%%%%%%%%%%%%%%%%%%%
%%%%%%%%%%%%% ACK %%%%%%%%%%%%%%%%%%%%%%%%%%%%%%%%%%%%%%%%%%%%%%%%%%%%%%%
%%%%%%%%%%%%%%%%%%%%%%%%%%%%%%%%%%%%%%%%%%%%%%%%%%%%%%%%%%%%%%%%%%%%%%%%%

\section{Acknowledgement}
\label{sec:ack}
This work has been supported by The French ANR research project MuTAS (ANR-21-CE22-0025-01).

%%


\bibliographystyle{IEEEtran}
\bibliography{main}
\end{document}